\documentclass[12pt]{iopart}
\usepackage{iopams}

\usepackage{amssymb}
\usepackage{graphicx}

\newcommand{\npart}{$N_{part}$ }

\newcommand{\detaabs}{$|\Delta\eta|$ }
\newcommand{\deta}{$\Delta\eta$ }
\newcommand{\dphi}{$\Delta\phi$ }
\newcommand{\dphiabs}{$|\Delta\phi|$ }
\newcommand{\detano}{$\Delta\eta$}
\newcommand{\dphino}{$\Delta\phi$}
\newcommand{\detaj}{$\Delta\eta(J)$ }
\newcommand{\dphij}{$\Delta\phi(J)$ }
\newcommand{\dphijr}{$\Delta\phi(J+R)$ }
\newcommand{\detajno}{$\Delta\eta(J)$}

\newcommand{\dphijrno}{$\Delta\phi(J+R)$}
\newcommand{\pttrig}{$p_{t}^{trig}$ }
\newcommand{\ptass}{$p_{t}^{assoc}$ }
\newcommand{\pttrigno}{$p_{t}^{trig}$}
\newcommand{\ptassno}{$p_{t}^{assoc}$}

\newcommand{\zt}{$z_T$ }
\newcommand{\mpt}{$p_t$ }

\begin{document}

\title{Intra-jet correlations of high-$p_t$ hadrons from STAR}
\author{J\"orn Putschke for the STAR Collaboration} 
\address{Lawrence Berkeley National Laboratory, Berkeley, California 94720, USA}

\ead{jhputschke@lbl.gov}

\begin{abstract}

Systematic measurements of pseudorapidity (\detano) and azimuthal
(\dphino) correlations between high-$p_t$ charged hadrons in
$\sqrt{s_{NN}}$=200 GeV Au+Au collisions are presented. An enhancement
of correlated yield at large \deta on the near-side is observed. This
effect persists up to trigger \pttrig $\sim$ 9 GeV/c, indicating that it is
associated with jet production. More detailed analysis suggests
distinct short-range and long-range components in the correlation.

\end{abstract}



Dihadron azimuthal correlation studies in nuclear collisions
have shown that hard partons interact strongly with the matter that is generated and
provide a sensitive probe of the medium \cite{star1,star2,star_highpTcoor}. Enhanced
near-side (\dphi $\sim$ 0) correlated yield at large \deta (the {\it ridge}) has been observed in
measurements with trigger particles at intermediate
\mpt (4 $<$ \pttrigno $<$ 6 GeV) \cite{jacobs,dan} and for dihadron pairs having $p_t$ $<$ 2 GeV
but no trigger requirement \cite{star_deta}. However, inclusive hadron
production at $p_t\lesssim 6$ GeV/c exhibits large differences between nuclear
collisions and more elementary collisions
\cite{star_LamK,star_piprot}. It is therefore unclear from these existing measurements whether
the ridge is associated with hard partonic scattering and jet
production. In this proceeding we extend the near-side correlation
measurement to \pttrigno$\sim9$ GeV/c, well into the kinematic region
where inclusive hadron production is similar in nuclear and elementary
collisions and where jet fragmentation is thought to dominate. We observe
the persistence of the ridge effect to the highest measured trigger
$p_t$, suggesting that it is indeed associated with jet production. We further characterize the ridge, to
gain insights into its origin.

\begin{figure}[t]
\begin{minipage}[t]{0.49 \textwidth}

\begin{center}
\includegraphics[width=\textwidth]{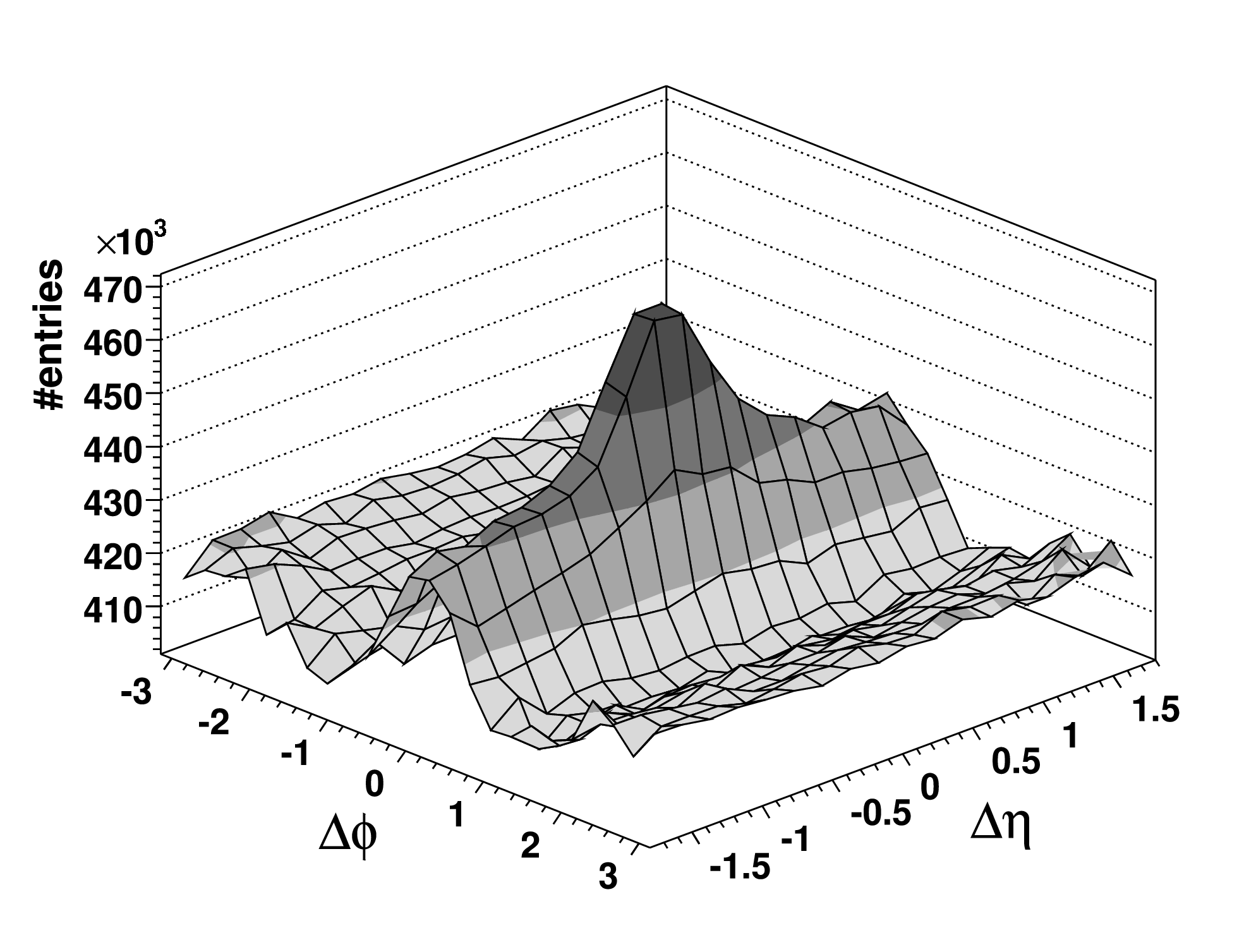}
\end{center}
\vskip -0.45cm
\caption{\label{pr}  Raw \deta $\times$ \dphi dihadron correlation function in central Au+Au collisions for 3 $<$\pttrigno$<$ 4 GeV and \ptassno$>$ 2 GeV.}

\end{minipage}\hfill
\begin{minipage}[t]{0.49 \textwidth}

\begin{center}
\includegraphics[width=\textwidth]{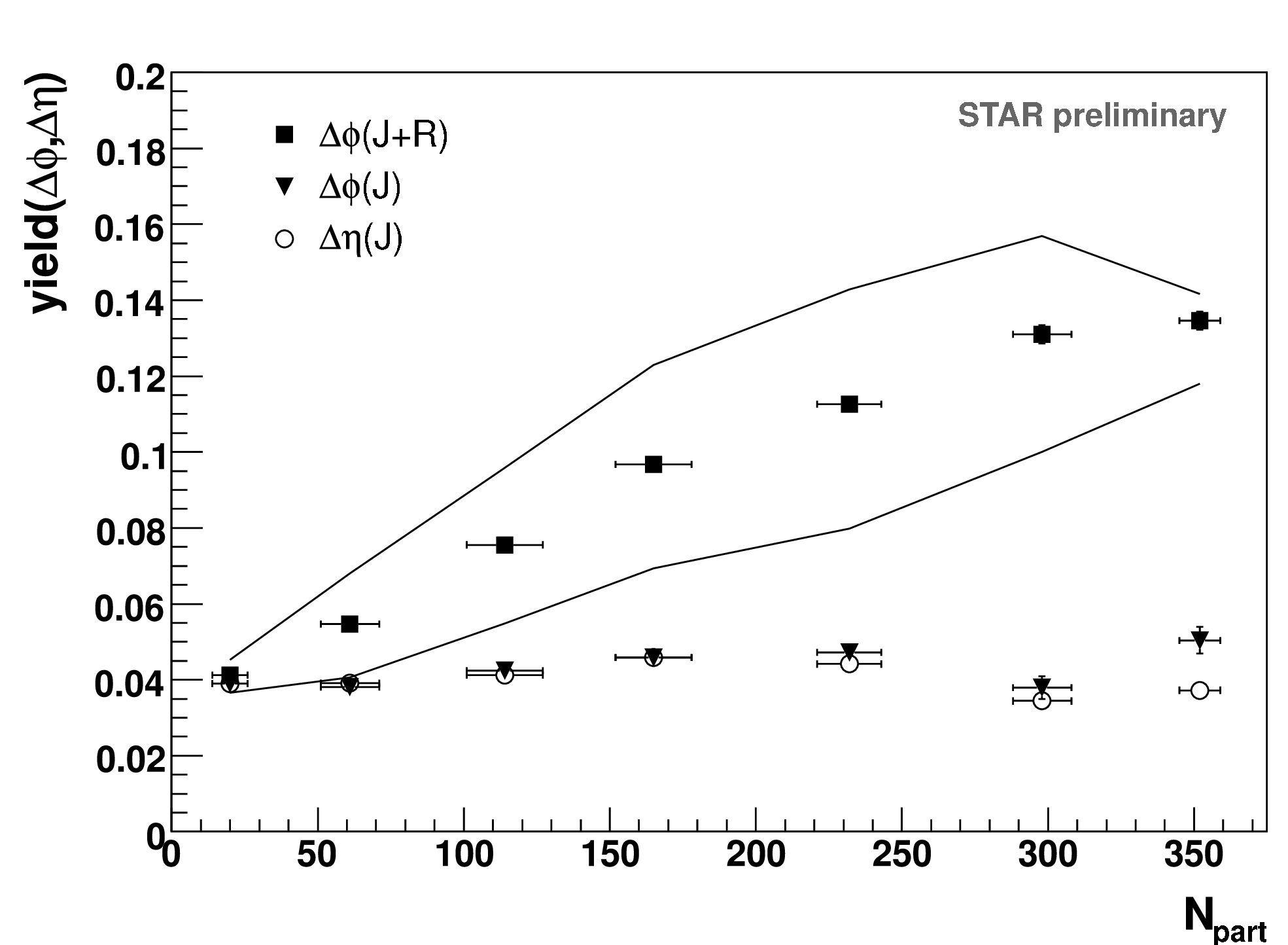}
\end{center}
\vskip -0.45cm
\caption{\label{fig1} Near-side yield of associated particles in \deta and \dphi with \ptass $>$ 2 GeV as a function of \npart in Au+Au for 3 $<$ \pttrig $<$ 4 GeV (details see text).}
\end{minipage}
\vskip -0.35cm 
\end{figure}

To illustrate the analysis method, Fig.\ \ref{pr} shows the \deta
$\times$ \dphi distribution of hadrons with \ptassno$>$ 2 GeV,
associated with trigger hadrons 3 $<$\pttrigno$<$ 4 GeV in central
Au+Au collisions. The yields are corrected for single-particle
tracking efficiency and the finite pair acceptance in \deta and
\dphino. The near-side yield shows a clear peak around (\detano,
\dphino) = (0,0), as expected from jet fragmentation. In addition a
prominent enhancement of correlated yield in \deta on the near-side is
clearly visible above the flow modulated background (the ridge). To
better understand the ridge phenomenon we decompose the near-side into
a \emph{jet}-like peak and a \deta independent \emph{ridge}
component. This \emph{ansatz} assumes distinct underlying physical
processes in the different \deta regions. We examine this assumption
below.

To extract the ridge yield from dihadron measurements we project the two dimensional (\deta $\times$ \dphino) correlation function  (Fig.\ \ref{pr}) onto \dphi and \deta in different \deta $\times$ \dphi regions. Three methods were used to characterize the small \deta \emph{jet-like} ($J$) and the large \deta \emph{ridge-like} ($R$) contributions to the near-side jet yield\footnote{The yields are extracted from bin-counting in the interval $|$\dphino $|$ $=$ $|$\deta $|$ $<$ 1} in \deta and \dphino:
\begin{itemize}
\item \dphijr: Projecting onto \dphi with the full experimental \deta acceptance ($|$\detano$|<1.7$ was used in this analysis) and subtracting the elliptic flow ($v_2$) modulated background\footnote{Not corrected for the finite \deta pair-acceptance.}. 
\item \dphij: Subtracting the \dphi projection for 0.7 $< $\detaabs $<$ 1.4 from the \dphi projection \detaabs $\leq$ 0.7 (near-side).
\item \detaj: Projecting onto \deta in a \dphi window \dphiabs $<$ 0.7 (near-side). A constant fit to the measurements was used to subtract the background. 
\end{itemize}
In Fig.\ \ref{fig1} the near-side yield is shown as a function of the number of participants \npart for all three methods. The agreement of the measured jet-like yield between the \detaj and \dphij method for all centrality bins, within the sensitivity of this analysis, supports the assumption that the ridge-like correlation is uniform in the \deta acceptance. Further detailed studies of the ridge shape in the high statistics Au+Au central data set, especially at high \pttrigno, will be pursued. Note that the jet-like correlated yield is independent of centrality and agrees with the p+p reference measurements \cite{dan}. In contrast the \dphijr yield shows a significant increase with centrality due to the inclusion of the correlated yield at large \deta (ridge). 

For the purpose of this analysis one can define the (absolute) $ridge$ $yield = yield($\dphijrno$)-yield($\detajno$)$\footnote{The ridge yield depends on the \deta integration window used in the \dphijr method.}. The main systematic error is the uncertainty in the elliptic flow measurement for the \dphijr method.  The $v_2$ value used in this analysis is the mean of the reaction plane ($v_2\{RP\}$) and four-particle cumulant method ($v_2\{4\}$) in Au+Au collisions \cite{starflow}. The systematic uncertainties were estimated using $v_2\{RP\}$ as maximum and $v_2\{4\}$ as minimum $v_2$ values (represented as lines in all figures). 

\begin{figure}[t]
\begin{minipage}[t]{0.49 \textwidth}

\begin{center}
\includegraphics[width=\textwidth]{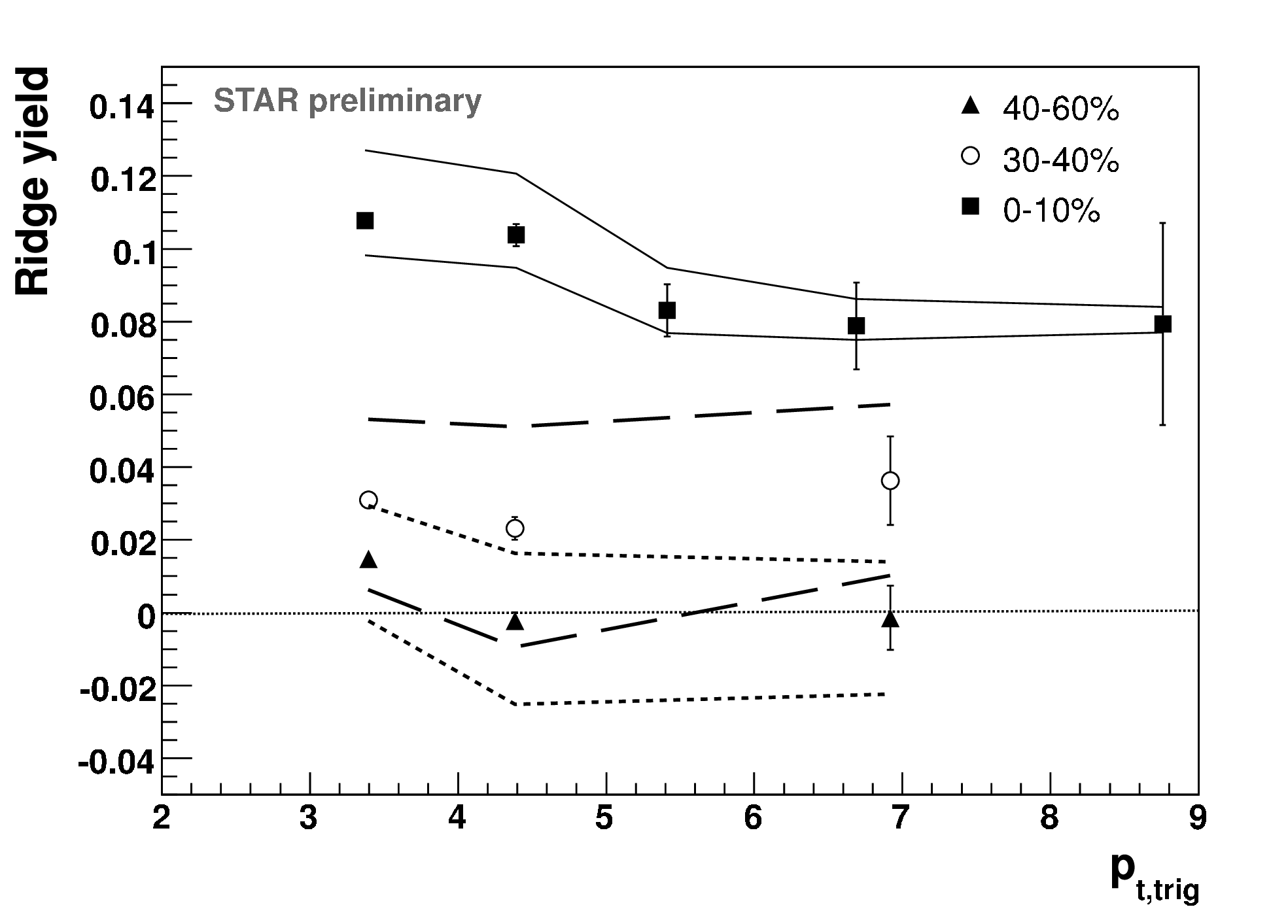}
\end{center}
\vskip -0.45cm
\caption{\label{fig3} (Absolute) Ridge yield for different centralities as a function of \pttrig for \ptass $>$ 2 GeV in Au+Au.} 

\end{minipage}\hfill
\begin{minipage}[t]{0.49 \textwidth}

\begin{center}
\includegraphics[width=\textwidth]{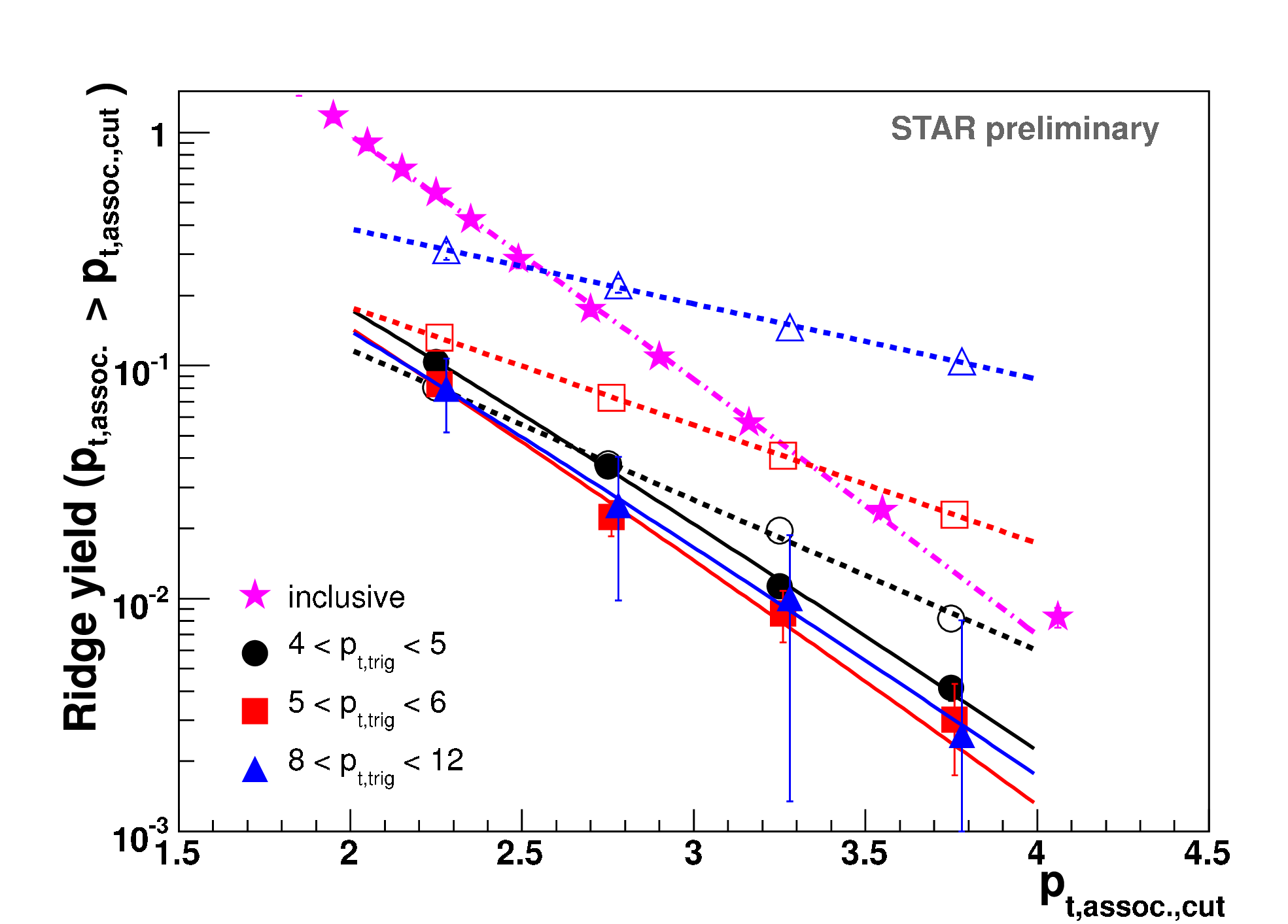}
\end{center}
\vskip -0.45cm
\caption{\label{fig5} (color online) Ridge/Jet-like yield (filled/open symbols) as function of \ptass  for different \pttrig in 0-10\% central Au+Au collisions. The inclusive spectrum (0-5\% central Au+Au, stars) is also shown \cite{star_pt}. } 

\end{minipage} 
\vskip -0.25cm
\end{figure}

Fig.\ \ref{fig3} shows that a significant (absolute) ridge yield
persists up to the highest \pttrigno, with yield increasing with
centrality. The finite ridge yield at \pttrig up to 9 GeV, where
parton fragmentation is expected to be the dominant hadron production
mechanism \cite{star_LamK,star_piprot}, indicates that the ridge is associated with
jet production.

To characterize in more detail the properties of particles associated
to the ridge-like or jet-like near-side correlation we use the \ptass
spectrum in different \pttrig windows, as shown in Fig.\
\ref{fig5}. An exponential function $\frac{dN}{dp_t}\propto p_t
e^{-p_t/T}$ is fitted to the data (lines in Fig.\
\ref{fig5}) to extract the inverse slope
parameter $T$. A clear difference between the slopes of the jet-like
yield, using the \detaj method, and the ridge-like yield is seen:
while the $p_t$ dependence of the ridge yield is similar to the
inclusive particle production, the jet-like associated yield has a
significantly harder $p_t$-spectrum, increasing with \pttrigno, as
expected from jet fragmentation.  The slope of the ridge-like yield is
largely independent of \pttrig and only slightly harder than the
inclusive spectrum with a slope difference $\Delta T \approx$ 40-50
MeV.

\begin{figure}[t]
\begin{minipage}[t]{0.49 \textwidth}

\begin{center}
\includegraphics[scale=0.4]{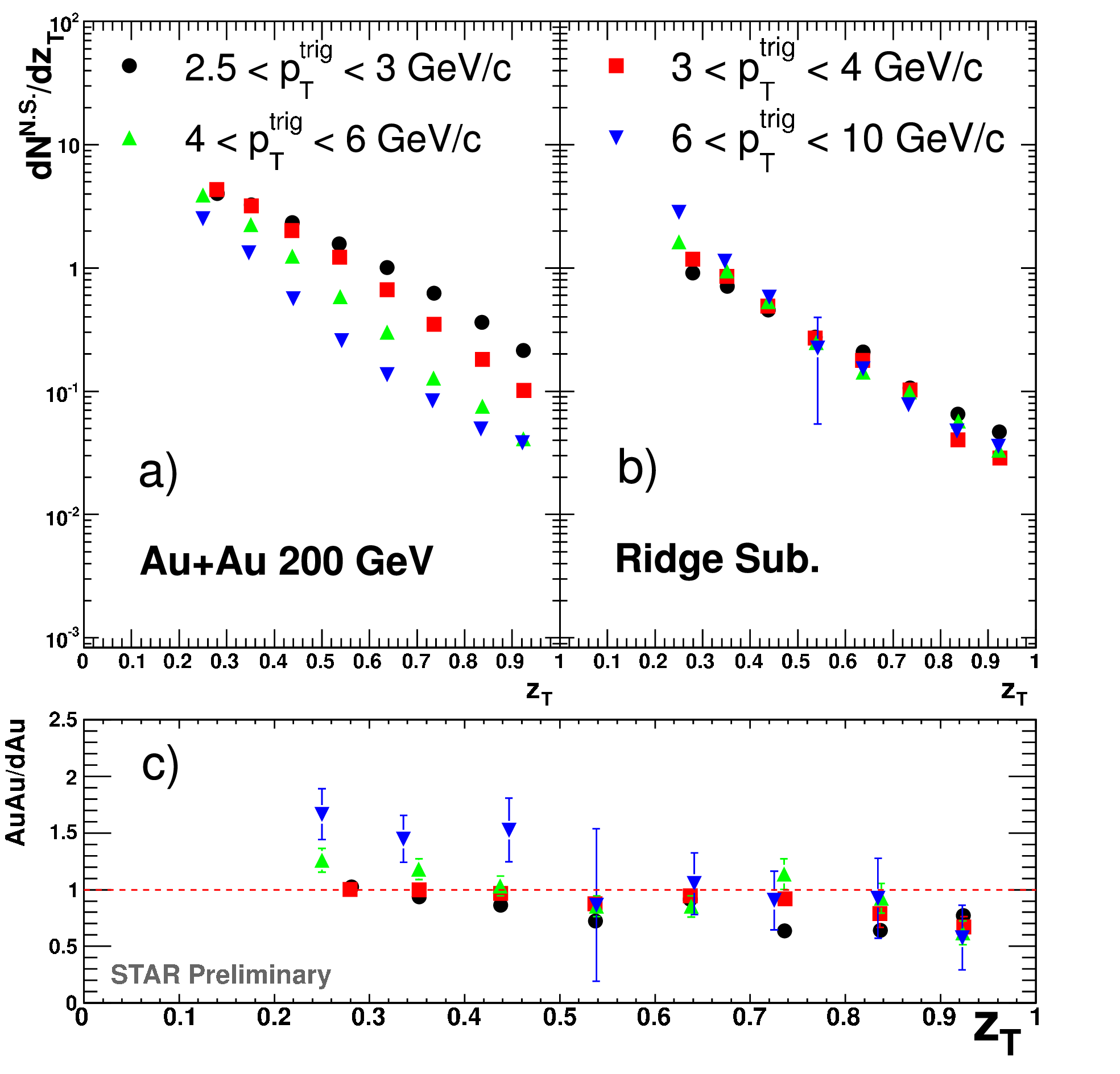}
\end{center}

\end{minipage}\hfill
\begin{minipage}[t]{0.49 \textwidth}
\caption{\label{zt} (color online) Near-side \zt di-hadron fragmentation function for different \pttrig in central Au+Au collisions before a) and after ridge subtraction b) as well as the ratio to d+Au reference measurements c) (see also \cite{mark}).}
\end{minipage} 
\vskip -0.45cm
\end{figure}

Fig.\ \ref{zt} a) shows the \pttrig dependence of the near-side \zt
di-hadron fragmentation function ($z_T=$\ptassno$/$\pttrigno) in
central Au+Au collisions (for details see \cite{mark}). Subtracting
the ridge-like contributions, using the \dphij method, one observes a
near-side fragmentation that is approximately independent of \pttrig
(Fig.\ \ref{zt} b)). The \zt distributions in central Au+Au collisions
after subtracting the ridge contribution are comparable to the d+Au
reference measurements (Fig.\ \ref{zt} c)) in contrast to the
non-ridge subtracted distributions \cite{mark}. To further quantify this
observation, studies to estimate the effect of background
fluctuations in the \dphij method at high \deta will be pursued.

These observations support the ansatz that the near-side \deta
$\times$ \dphi correlation consists of two distinct components: a jet
contribution, consistent with the p+p and d+Au dihadron reference
measurements \cite{jacobs,dan}, and the ridge contribution with
properties similar to the medium. This could arise from partonic
energy loss followed by fragmentation in vacuum, with the lost energy
appearing dominantly in the ridge.

Several models are qualitatively able to describe the presented
phenomena: coupling of induced radiation to longitudinal flow
\cite{Armesto_flow}, turbulent color fields \cite{Majumder}, anisotropic plasma \cite{Romatschke}, a combination of jet-quenching and strong radial
flow \cite{Voloshin_flow} and recombination of locally thermal
enhanced partons due to partonic energy loss in the recombination
framework \cite{Hwa_flow}. A comparison of quantitative theoretical
calculations to the measurements are needed in order to understand the
origin of the ridge.

\section*{References}

\bibliographystyle{epj}
\bibliography{ref}

\end{document}